\documentclass[aps,pre,twocolumn,showpacs]{revtex4}
\usepackage{graphicx}
\usepackage{amsmath}

\begin{document}

\title{Fast fixation without fast networks}

\author{Gareth J.\ Baxter$^{1}$, Richard A.\ Blythe$^{2}$, and 
Alan J.\ McKane$^{3}$}
\affiliation{$^{1}$Department of Physics \& I3N, University of Aveiro, 
3810-193 Aveiro, Portugal \\
$^{2}$SUPA, School of Physics and Astronomy, University of Edinburgh,
Mayfield Road, Edinburgh EH9 3JZ, UK \\
$^{3}$Theoretical Physics Division, School of Physics and Astronomy, 
University of Manchester, Manchester M13 9PL, UK}
 
\begin{abstract}
We investigate the dynamics of a broad class of stochastic copying processes on 
a network that includes examples from population genetics (spatially-structured 
Wright-Fisher models), ecology (Hubbell-type models), linguistics (the utterance
selection model) and opinion dynamics (the voter model) as 
special cases. These models all have absorbing states of fixation
where all the nodes are
in the same state. Earlier studies of these models showed that the mean time 
when this occurs can be made to grow as different powers of the
network size by varying 
the the degree distribution of the network. Here we demonstrate that this 
effect can also arise if one varies the asymmetry of the copying dynamics 
whilst holding the degree distribution constant. In particular, we show that 
the mean time to fixation can be accelerated even on homogeneous networks when 
certain nodes are very much more likely to be copied from than copied to. We 
further show that there is a complex interplay between degree distribution and 
asymmetry when they may co-vary; and that the results are robust to 
correlations in the network or the initial condition.
\end{abstract} 

\pacs{05.40.-a,89.75.Hc,87.23.-n,87.23.Ge}

\maketitle

\section{Introduction}
\label{sec:intro}
One of the central themes in the application of the ideas and techniques of
non-equilibrium statistical physics to the modeling of biological and 
social systems, is that of agents interacting through a network of links 
\cite{Dorogovtsev03,Newman10}. The agents may be individuals, species, 
companies, or other kinds of entity, and the nodes of the network may consist 
of one or many agents, but the general idea is the same. An agent at one node 
$i$ interacts with another at node $j$ if a link joining the two nodes is 
present. The probability of interaction may depend on the strength of the link 
or on the properties of the agents themselves. This stochastic dynamics may 
also include the birth or death of agents, their transformation from one type 
to another, or other more complicated processes.

In very many applications, the network structure is defined through a single 
symmetric matrix $G_{ij}$, whose entries give the strength of the link joining 
node $i$ to node $j$. Quantitatively, this strength might specify the frequency
that the two agents at sites $i$ and $j$ come together to interact.  Most 
simply, the entries may be zero if the link is absent and one  if it is 
present: $G$ is then the adjacency matrix for the network. However in some 
systems, particularly in the social sciences, even variation in link strength 
or interaction frequency is not the whole story. Individuals may interact 
strongly or weakly, frequently or infrequently, and the nature of the  
interaction may, for  instance, be antagonistic, neutral or reinforcing, or one 
of the agents may  have significantly more impact than the other. To model  
these aspects, one may define another matrix, $H_{ij}$, which quantifies the 
nature of the influence that an agent at node $j$ has on one at node $i$. A key 
property of the $H$ matrix that distinguishes it from $G$ is that it need not 
be symmetric: agent $i$ may have much more influence on agent $j$ than vice 
versa.

This decomposition of interactions into symmetric and asymmetric parts turns out
to be extremely natural in the case of the utterance selection model
for language
change that we introduced a number of years ago \cite{Baxter06}.  In this 
model, the nodes of the network represent speakers who have the possibility 
of saying the same thing in two (or more) different ways.  The process
of language
change is assumed to be the consequence of repeated face-to-face interactions
between speakers, and so the frequency $G_{ij}$ that the pair of individuals
$(i,j)$ interacts must necessarily be symmetric. However the weight that
individual $i$ gives to the utterances of individual $j$ may depend on factors
other than the frequency of interaction, such as the relative social standing.  
Whilst the frequency that $i$ interacts with $j$ must necessarily equal 
the frequency that $j$ interacts with $i$, there is no reason why the 
agents should judge each other to be of similar social standing. Such 
asymmetric effects can enter only via the matrix $H_{ij}$.

In this work, we systematically investigate the effect that varying
the asymmetry (the matrix $H$) has on the dynamics of the utterance
selection model.  The basic microscopic process at work in this model is one
agent replicating the behavior that another agent has previously exhibited.
As such, the utterance selection model is a member of a much larger class of 
stochastic copying processes. Other models within this class include the
Wright-Fisher model for changes in gene frequencies in a population
\cite{Fisher30,Wright31,Crow70,Ewens04,Barton07}, Hubbell's model for species
diversity in an ecological community \cite{Hubbell2001} and the voter model
that has been widely studied by statistical physicists as a baseline model
of opinion dynamics \cite{Castellano2009,Sood05,Suchecki05,Suchecki05b,Sood08}.

We remark that in many of these cases, the
asymmetry encoded in the matrix $H$ is a side-effect of the microscopic update
rule that defines the model, as opposed to a quantity that can be varied
independently in its own right. For example, the voter model is defined in terms
of the following update: first, a site of the network is chosen at random,
and then the state of that site is updated to match that of a
randomly-chosen neighbor. This choice of update rule then implies that
well-connected nodes are much more influential than  poorly-connected
nodes, as they are more likely to be copied from than copied to.
Nevertheless, the network structure of such a model (the matrix $G$)
is easily varied, and by doing so it has been found that the mean time
to reach a state where all nodes have the same state---variously known
as \emph{fixation}, 
\emph{consensus} or \emph{complete order}---can grow as different powers of 
the network size $N$ \cite{Sood05,Suchecki05,Suchecki05b,Sood08} depending on 
the level of heterogeneity in the network structure.

By exploiting the clean separation of network structure $G$ and interaction
asymmetry $H$ afforded by the utterance selection model, we show that, even when
the network structure is homogeneous, disparities in the impact of
different agents,
as expressed through the $H$ matrix, may drive the system more quickly to
fixation than when such disparities are absent. Thus we may arrive at
fast fixation
without the need for special `fast' network structures, as observed in
previous works,
if we instead manipulate the asymmetry in the interactions. One can of
course also
consider the case where network structure ($G$) and asymmetry ($H$) co-vary.
As we discuss later in this work, this leads to a wide variety of
scaling relations between
the network size and the mean time to reach fixation.

This work builds on our earlier investigations of the utterance
selection model, in which we
introduced the model and studied the case of a fully-connected network with
a constant $H_{ij}$~\cite{Baxter06}, investigated its application to the 
emergence of New Zealand English~\cite{Baxter09}, and studied the effect of 
the network structure on the mean time to fixation~\cite{Baxter08}. In the 
latter paper we considered the model in a broader context, which included 
models of population ecology and population genetics, where the $G$ and $H$ 
matrices appeared together in a migration matrix $m_{ij} \equiv G_{ij}H_{ij}$. 
We showed that if $m_{ij}$ was symmetric, then the mean time to fixation was
essentially independent of the network structure. Since $G_{ij}$ is 
symmetric, fixation times much shorter or longer than this can only be 
found if $H_{ij}$ is not symmetric. However, in Ref.~\cite{Baxter09}, we found
that short fixation times would be needed to explain the rapid 
emergence of New Zealand English. We therefore postulated that there must
have been a fraction of individuals in the population who had, for instance, 
greater influence than average, leading to a skewed distributed for $H_{ij}$,
giving a let-out from the results of Ref.~\cite{Baxter08}.  The effect
of these skewed distributions on the mean time to fixation form the
focus of this present work.

The outline of the paper is as follows. In Sec.~II we define the model and 
further develop the formalism that we will use in the rest of the paper. In 
Sec.~III and IV this is applied to investigate how the structure of the 
matrices $G_{ij}$ and $H_{ij}$ influence the long time dynamics of the model.
We consider two distinct cases: one in which the influence encoded in
$H_{ij}$ is independent of the network structure described by
$G_{ij}$, and another in which they are directly
related to one another.  In the former case we find that influence may
accelerate the fixation process; while in the latter we find a wide
variety of behavior that is summarized in Fig.~\ref{exponent_chart}. We
conclude in Sec.~V with a summary of our findings and how they relate
to studies of similar models. An Appendix contains some useful
mathematical results that are employed in Secs.~II and III.

\section{Model and Formalism}
\label{sec:model}
The system, when expressed in terms of the model of language change 
mentioned in the Introduction~\cite{Baxter06}, consists of $N$ speakers, 
whose frequency of interaction is given by a matrix $G$. More specifically, 
speakers $i$ and $j$ interact with a probability $G_{ij}$, normalized so 
that $\sum_{\langle ij \rangle}G_{ij}=1$,
where $\langle ij \rangle$ refers to distinct pairs $i$ and $j$. In this simple
version of the model, we will only monitor the frequency with which two 
different ways of saying the same thing spreads through the speaker 
community. That is, as in \cite{Baxter06}, we will focus only on a
single expression with two variants, or linguemes, which we denote as
$a$ and $b$. 

The state of the system is completely specified by the probabilities
for each speaker to utter the $a$ variant at a given time $t$. These will
be denoted by $x_{i}(t)$; the rule by which they are determined is given
below. We will frequently express the overall state of the system as
$\mathbf{x}\equiv(x_1,\ldots,x_{N})$. The second matrix mentioned in the 
Introduction, $H$, specifies how much weight $i$ gives to the utterances of 
$j$. With the structure of the model in place, it remains for the dynamics 
to be specified. The evolution of language use will be taken to be 
usage-based \cite{Croft00,Bybee01,Tomasello03}:
a speaker will be influenced by the extent to which a particular variant
is used by the speaker he is in conversation with.
In the formulation introduced in~\cite{Baxter06}, it was assumed that a
conversation would consist of $T$ tokens, i.e.~instances of use, of the  
particular word or expression that is of interest. Here we will 
simply take $T=1$. This choice will not significantly change the nature of
the dynamics, and moreover the choice of $T$ amounts to a rescaling of $G$
and $H$, and so it may be reintroduced at any time by performing these 
rescalings.

The actual production process is expected to be stochastic~\cite{Croft00},
and therefore we adopt the rule that at time $t$ speaker $i$ produces 
variant $a$ with probability $x_i(t)$ and variant $b$ with
probability $(1-x_i(t))$. This is represented by a stochastic variable
$\zeta_i$, so that $\zeta_i = 1$ with probability $x_{i}(t)$ and zero
otherwise. The change in the grammar of speaker $i$ due to an interaction 
(conversation) with speaker $j$ will be of the form
$(\zeta_{i} + H_{ij}\zeta_{j})$.  The first term is the result of speaker $i$ 
uttering an $a$ variant ($\zeta_{i}=1$) or not (i.e. uttering a $b$ variant, 
$\zeta_{i}=0$) and the second the result of speaker 
$j$ uttering an $a$ variant ($\zeta_{j}=1$) or not ($\zeta_{j}=0$). The
weight given to the utterance of $j$ by $i$ is the factor $H_{ij}$ discussed
above.

There are two further factors that have to be introduced. First, we multiply 
the above interaction term by a constant $\lambda$, which is taken to be
small, since grammatical changes as a result of a single conversation will
typically be small. Second, we have decided to choose the random variable 
$\zeta$ to be one or zero, following the original model~\cite{Baxter06}. Using 
our convention, the overall value of $x_i$ has to be renormalized 
(by a factor of $[1+\lambda(1+H_{ij})]$) at each update. An alternative choice 
would be to take $\zeta$ to be one or minus one, which would avoid the need 
to normalize. With an appropriate correction to the values of $H_{ij}$ the 
two choices are equivalent.

If we now assume that one conversation takes place during a time $\delta t$,
and that this conversation has been between the two speakers $i$ and $j$, 
then the change in the grammar of speaker $i$ as a result of this interaction
will be
\begin{equation}
x_i(t+\delta t) = \frac{x_i(t)+\lambda(\zeta_{i} + H_{ij}\zeta_{j})}
{1+\lambda(1+H_{ij})},
\label{update_rule}
\end{equation}
with a similar equation for speaker $j$ obtained by interchanging the indices 
$i$ and $j$. 

An alternative mechanistic description consists of viewing speaker $i$
as containing a large number of objects, of which a fraction $x_i(t)$
are of type $a$ and a fraction $(1-x_i(t))$ of type $b$.  One object
is then picked at random for ``migration'' from speaker $i$ to another
speaker. This was the formulation discussed in~\cite{Baxter08}:
speakers were viewed as islands containing individuals of a species
which could undergo birth/death and
migration. This picture shows how other models such as the
Wright-Fisher population genetics model
\cite{Fisher30,Wright31,Crow70,Ewens04,Barton07}  or the Hubbell
ecology model \cite{Hubbell2001} may be treated with the same
formalism we have outlined here. The relationships between these
models are discussed in more detail in \cite{Blythe2007b}.

In simulations we repeatedly use the update rule 
(\ref{update_rule}), after choosing the two speakers who are interacting 
using the network structure matrix $G_{ij}$. However, to make analytic 
progress we take $\delta t \to 0$, and construct a Fokker-Planck equation 
for the Markov process (\ref{update_rule}). The derivation is given 
in~\cite{Baxter06}, where it is shown that the probability of the system 
being in state $\mathbf{x}$ at time $t$, $P(\mathbf{x},t)$, satisfies the
equation
\begin{eqnarray}
& & \frac{\partial P}{\partial t} = \sum_{\langle ij \rangle} 
\left( m_{ij} 
\frac{\partial }{\partial x_i} - m_{ji} \frac{\partial }{\partial x_j} \right)
\left[ \left( x_i - x_j \right) P \right] \nonumber \\
& & +\frac{1}{2} \sum_{i=1}^{N} G_{i} \frac{\partial^2 }{\partial x^2_i} 
\left[ x_i\left( 1-x_i \right) P \right],
\label{FPE}
\end{eqnarray}
where $G_i \equiv \sum_{\langle ij \rangle} G_{ij}$ and 
$m_{ij} \equiv G_{ij}h_{ij}$. Here $h_{ij}$ is $H_{ij}$ rescaled in a way 
that is appropriate for the Fokker-Planck description of the model. The 
precise relationship between them is $H_{ij} = \lambda h_{ij}$, and since by
construction $h_{ij}$ must be independent of $\lambda$, when we use $H_{ij}$ in
the context of a Fokker-Planck description, it is to be understood as being
proportional to $\lambda$.

The Fokker-Planck equation (\ref{FPE}) seems far too complicated to be amenable 
to analysis, but remarkably progress can made \cite{Baxter08,Blythe2010}.
The reason for this lies in
the fact that after a relatively short time (compared to the very long fixation 
times that are of interest to us here) the change in the speakers grammars
effectively become coupled, and their dynamics can be described by a single 
collective variable
\begin{equation}
\xi(t) = \sum^{N}_{i=1} Q_i x_i(t),
\label{xi}
\end{equation}
where $Q_{i}$ will be defined below. The problem now reduces to one having a 
single degree of freedom. Methods based on the backward Fokker-Planck
equation~\cite{Gardiner04,Risken89} can then be used to obtain an expression 
for the mean time to fixation. Precise criteria for determining the validity 
of this reduction to a single coordinate are given in \cite{Blythe2010}.
Here we content ourselves with the observation that these criteria are usually 
satisfied when the network has sufficiently small diameter, and by checking 
our analytical predictions against Monte Carlo simulations.

To define $Q_i$ we follow \cite{Baxter08} and introduce a matrix $M_{ij}$ by
\begin{equation}
M_{ij} = \left\{ \begin{array}{ll} 
m_{ij}, & \mbox{\ if $j \neq i$} \\ 
- \sum_{k \neq i} m_{ik}, & \mbox{\ if $j = i$}.
\end{array} \right.
\label{M}
\end{equation}   
From this it follows that $\sum_{j} M_{ij} = 0$, that is, $M_{ij}$ has at 
least one eigenvalue equal to zero (assumed unique) with the corresponding 
right-eigenvector having all elements equal to one. The corresponding 
left-eigenvector (suitably normalized) defines $Q_i$: 
\begin{equation}
\sum^{N}_{i=1} Q_{i}M_{ij} = 0 \ \ {\rm with} \ \ \sum^{N}_{i=1} Q_{i} = 1.
\label{Q}
\end{equation}
We can make some interesting observations regarding the dynamics of the mean
of $\xi(t)$, by first noting that from the Fokker-Planck equation (\ref{FPE}) 
the mean of $x_i(t)$ evolves according to
\begin{equation}
\frac{{\rm d}\langle x_{i}(t) \rangle}{{\rm d}t} = \sum_{j \neq i} m_{ij} 
\left( \langle x_{j}(t) \rangle - \langle x_{i}(t) \rangle \right) \equiv
\sum^{N}_{j=1} M_{ij} \langle x_{j}(t) \rangle. 
\label{evolution_x}
\end{equation}
This implies that the average of $\xi(t)$ is conserved by the dynamics:
\begin{equation}
\frac{{\rm d}\langle \xi(t) \rangle}{{\rm d}t} = 
\sum_{i=1}^{N} Q_{i} \frac{{\rm d}\langle x_{i}(t) \rangle}{{\rm d}t} 
= \sum_{i,j} Q_{i}M_{ij} \langle x_{j}(t) \rangle = 0,
\label{evolution_xi} 
\end{equation}
where we have used Eq.~(\ref{Q}). A solution of Eq.~(\ref{evolution_x}) gives
$\langle x_i(t) \rangle$ as an expansion in terms of the right eigenvectors 
of $M$. In the $t \to \infty$ limit only the one corresponding to the zero
eigenvalue survives, but we have seen that all the elements of this 
particular eigenvector are equal. So 
$\lim_{t \to \infty} \langle x_i(t) \rangle$ is independent of $i$. Since all 
$x_{i} (t)$ tend to $0$ or $1$ as $t \to \infty$, this is the probability of 
the variant $a$ fixing. Taking the average of Eq.~(\ref{xi}), letting 
$t \to \infty$, and using $\sum_{i} Q_i = 1$, we see that this is also the 
value of $\lim_{t \to \infty} \langle \xi(t) \rangle$. So the fixation 
probability
is $\langle \xi(\infty) \rangle$. However, from Eq.~(\ref{evolution_xi})
we recall that $\langle \xi(t) \rangle$ is conserved, so the fixation 
probability is also $\xi (0)$.

These are straightforward deductions that we can make simply by considering
the mean values of $x_i(t)$ and $\xi(t)$. To make further progress one has
to solve the backward Fokker-Planck equation as indicated above. This is 
carried out in \cite{Baxter08}, where it is shown that, under reasonable 
assumptions that are expanded on in \cite{Blythe2010}, the mean time to
fixation is given by
\begin{equation}
T = - \frac{2}{r} \left[ \xi(0) \ln \xi(0) + (1-\xi(0))\ln(1-\xi(0))\right],
\label{fix_time}
\end{equation}
where
\begin{equation}
r \approx \sum_{i} Q^{2}_{i} G_{i} \frac{2\sum_{j \neq i} m_{ij}}
{2\sum_{j \neq i} m_{ij} + G_{i}}.
\label{r}
\end{equation}
So, in principle, we can find the mean fixation time from a knowledge of
the matrices $G$, and $H$ and the vector $Q$. The first two are assumed 
given---they characterize the system under consideration. Only $Q$, the left 
eigenvector of $M$ corresponding to zero eigenvalue, needs to be found. 

The next section of the paper will be devoted to an analytical study of this
question for various choices of the matrix $H$, and the subsequent section to
a numerical study. This latter section will both explore choices which cannot 
be treated analytically and will also be used to check the validity of the 
various approximations that are made in the derivations presented in the paper.
However, let us end this section by recalling the case where the analysis is 
most straightforward~\cite{Baxter08}. If $m$ is symmetric (and so $H$ is 
symmetric, since $G$ always is), then the right and left eigenvectors of $M$ 
must be identical. Therefore $Q_i$ must be the same for all $i$ and so from 
the normalization condition $Q_{i}=1/N$. The constant $r$ is now given in 
terms of known quantities. Incidentally, in this case the interpretation of 
$\xi(t)$ is especially clear --- as a ``center-of-mass'' coordinate: 
\begin{equation}
\xi (t) = \frac{1}{N} \sum^{N}_{i=1} x_{i} (t).
\label{CoM}
\end{equation}
The object of this paper is to investigate mean fixation times when $H$
is not symmetric, that is, when the relationship between speakers is not 
symmetric. This is what we now turn to.

\section{Analytic calculations of mean fixation time}
\label{sec:analytic}
We have seen that the case where $H_{ij}$ is symmetric leads to a $Q_i$ which 
is equal to $1/N$ for all $i$. If we go further and ask that $H_{ij}$ is a 
constant (i.e.~independent of $i$ and $j$) then we can also show that 
the right-hand side of Eq.~(\ref{r}), and so the mean time to fixation, is 
independent of the network structure \cite{Baxter08}. To go beyond this and 
make analytical progress we have to make specific assumptions for the form 
of $H_{ij}, G_{ij}$ or both.

One particular form for $H_{ij}$ which allows us to make such progress, is to 
assume that $H_{ij}$ is separable: $H_{ij}=\alpha_{i}\beta_{j}$. This is not
an unreasonable assumption; it allows us to look at the case where speakers 
are influenced by (the $\alpha_i$) or influence (the $\beta_j$) other speakers
irrespective of the identity of their interlocutor.  

Under this assumption, the solution for $Q_i$ becomes simple:
\begin{equation}
Q_i = \frac{\beta_i/\alpha_i}{\sum_j \beta_j/\alpha_j}\;.
\label{Q_i}
\end{equation}
It is straightforward to verify that this is a left-eigenfunction of $M$ 
with zero eigenvalue. This result can be understood by interpreting the matrix 
element $M_{ij}$ as the rate at which a particle hops from site $i$ to site
$j$ of the network.  Application of a Kolmogorov criterion \cite{Kelly79} then 
reveals that the separable form of $H_{ij}$ implies that detailed balance is 
satisfied, i.e., that $Q_i M_{ij} = Q_j M_{ji}$. Then (\ref{Q_i}) is the unique 
normalized solution of this set of equations, and we can write $r$ explicitly as
\begin{equation}
r = \frac{1}{[\sum_j\beta_j/\alpha_j]^2} 
\sum_{i} \frac{\beta^{2}_{i}}{\alpha_i^2} G_{i} \frac{2\alpha_i\sum_{j \neq i}
  G_{ij}\beta_{j}}
{2\alpha_i\sum_{j \neq i} G_{ij}\beta_{j} + G_{i}}\;.
\label{explicit_r}
\end{equation}

The fixation time is proportional to $1/r$, and so we will focus on the 
calculation of $r$. Note, however, that $\xi(0)$ will depend on the initial 
values of $x$. This is turn may have a (relatively weak) effect on the 
fixation time. Here we will assume that $x_i(0) = x_0\, \forall i$, then 
$\xi(0) = x_0$. 

A second assumption which allows analytic progress to be made is that the 
network of speaker interactions is large, random, and uncorrelated. It is then
defined by the degrees of the nodes, that is of the speakers, and we write
$G_{ij} \propto k_{i} k_{j}$, where $k_i$ is the degree of node $i$. Since the 
mean node degree, $\mu_1$, is given by $N^{-1}\sum_{i} k_i$ and 
$\sum_{i,j}G_{ij}=2$, the constant of proportionality is $2/(N\mu_1)^2$, and 
so we have
\begin{equation}
G_{ij} \approx \frac{2k_ik_j}{(N\mu_1)^2} \qquad \mbox{ and } \qquad G_i
\approx \frac{2k_i}{N\mu_1}\;.
\label{G_factor}
\end{equation}

If we assume both the decomposition of $H_{ij}$ and Eq.~(\ref{G_factor}) we 
obtain
\begin{equation}
r = \frac{1}{N\mu_1[\sum_j\beta_j/\alpha_j]^2} 
\sum_{i} \frac{{4\beta_i^2k_i}\sum_{j \neq i} k_{j}\beta_{j}}
{{2\alpha_i^2}\sum_{j \neq i} k_{j}\beta_{j} + \alpha_i(N\mu_1)}\;.
\label{assume_both}
\end{equation}

Under these two approximations, we can try out different schemes for
the interaction weightings. We are mainly interested in how the
fixation time scales with $N$, and are in particular looking for significant
deviations from the baseline result (found when $H_{ij}$ is a 
constant)~\cite{Baxter08} that $T$ is proportional to $N^2$.

\subsection{Asymmetry independent of network structure}
\label{sec:independent}

We first investigate the situation in which $H_{ij}$ is \emph{not} a function 
of degree, and hence $H_{ij}$ and $G_{ij}$ are statistically independent 
quantities. We will also assume that the $\alpha_i$ are all equal: $\alpha_i=1$,
say, while the $\beta_i$ take on arbitrary values. This means that different 
speakers' utterances carry different weights with their audience, but the 
importance given to them does not vary from listener to listener. Then
\begin{equation}
r = \frac{1}{N\mu_1[\sum_j\beta_j]^2} 
\sum_{i}\frac{{4\beta_i^2k_i}\sum_{j \neq i} k_{j}\beta_{j}}
{{2}\sum_{j \neq i} k_{j}\beta_{j} + N\mu_1}\;.
\label{r_fixed_alpha}
\end{equation}
Suppose that the $\beta_i$ are selected from some distribution. Since they are
selected independently from the $k_i$,
\begin{equation}
\sum_{j\neq i} k_j\beta_j \approx N\mu_1\langle
\beta\rangle - k_i\beta_i\;.
\label{j_notequal_i}
\end{equation}

There is now only the sum on $i$ remaining in Eq.~(\ref{r_fixed_alpha}). It
may be written in the form 
\begin{equation}
\sum_{i} \frac{\beta^2_i k_i \left[ 1-\delta_i \right]}
{\left[ 1-\epsilon_i \right]}\;,
\label{sum_on_i}
\end{equation}
where $\delta_i$ and $\epsilon_i$ are proportional to $k_i\beta_i/N$. So for
large $N$, we may expand the summand in Eq.~(\ref{sum_on_i}) in powers of 
$k_i\beta_i/N$ to obtain
\begin{eqnarray}
r &\approx&\frac{4}{N^2\mu_1\langle \beta\rangle^2}
\left\{\frac{\mu_1\langle\beta\rangle\langle\beta^2\rangle}
{[1+2\langle\beta\rangle]} - \frac{\mu_2\langle\beta^3\rangle}
{N\mu_1[1+2\langle\beta\rangle]^2} \right. \nonumber \\
&-& \left. \frac{2\mu_3\langle\beta^4\rangle}
{(N\mu_1)^2[1+2\langle\beta\rangle]^3} - \dots \right\}\;,
\label{r_beta_distn}
\end{eqnarray}
where $\mu_n$ is the $n^{\rm th}$ moment of the degree distribution.

If the $\beta_i$ are selected from a generic distribution, such as a Gaussian, 
the moments are well behaved, that is, they tend to a finite value for 
$N \to \infty$. This implies that $r \propto N^{-2}$ for large $N$ and so the 
mean time to fixation grows as $N^2$ for large $N$. This is identical to that 
obtained from the simplest case where $H_{ij}$ had no structure at all, and
suggests that if we are to look for deviations from this behavior then we
must investigate distributions where the moments depend on $N$ in some way.
One case in which this occurs is in `heavy-tailed distributions', which 
would correspond to our intuition that deviations from the $N^2$ behavior 
for the mean time to fixation might occur when there are members of the 
community who have a much larger influence than the modal value. If we
assume that the heavy tail has the structure of a power law, then we can make 
analytic progress, as discussed in the Appendix.

Returning to Eq.~(\ref{r_beta_distn}), we choose the distribution to be a power
law over its entire range, i.e.,
\begin{equation}\label{power_law_distn}
P(\beta) = A\beta^{-\gamma} \mbox{ for } \beta \geq \beta_0\;,
\end{equation}
and examine the dependence of $r$ on $N$ for different values of the exponent
$\gamma$ using Eq.~(\ref{powerlaw_scaling}) of the Appendix.
For instance, when $1<\gamma<2$, the ratio of the $N$-dependence of the three
terms in the large brackets in Eq.~(\ref{r_beta_distn}) is 
$N^{(3-\gamma)/(\gamma-1)}$ : $N^{1/(\gamma-1)}$ : $N^{1/(\gamma-1)}$,
and so the first term
dominates. For $2<\gamma<3$, the first moment $\langle\beta\rangle$ is a
constant, but a similar analysis shows that again the first term dominates. 
Finally, when $\gamma>3$, higher moments may also have a finite limit as 
$N \to \infty$, but once again it is found that the first term is the most 
important for large $N$. Therefore for a heavy-tailed distribution of this 
kind
\begin{equation}
r \approx \frac{4\langle\beta^2\rangle}{N^2\langle \beta\rangle
[1+2\langle\beta\rangle]}\;
\label{r_largeN}
\end{equation}
for large $N$. 

Since for $\gamma>3$ both $\langle\beta\rangle$ and $\langle\beta^2\rangle$ 
have finite limits as $N \to \infty$, we recover the $T\propto N^{2}$
result found from more conventional distributions. For $1<\gamma<2$, 
Eq.~(\ref{r_largeN}) gives $T\propto N$ and for $2<\gamma<3$, 
$T\propto N^{(5-3\gamma)/(\gamma-1)}$, and so in this range the power of $N$ 
varies from $3/2$ to $2$, having the former value when $\gamma=2$. So, in 
summary, choosing an extreme distribution for $\beta_i$ of the 
type (\ref{power_law_distn}) can reduce the growth of $T$ with population
size, the slowest growth (and hence the shortest fixation times) being 
for $\gamma \leq 2$ when $T \propto N$.

The complementary situation to the one we have just examined is to take the
$\beta_i$ to be all equal, while the $\alpha_i$ are free to vary. In this
situation, some speakers give more attention to others' utterances, and some 
less, but the identity of their interlocutor is not taken into account. 
However in this situation the method we used when the $\alpha_i$ were all
equal does not apply, and we have been unable to obtain any simple analytic 
results. We did carry out numerical simulations of this case, which
are detailed in Section \ref{sec:numerical} below.

\subsection{Asymmetry depends on speakers degree}
\label{sec:dependent}

A more extreme situation might be engineered by considering that a speaker's 
influence depends on the number of their interlocutors. This might be realistic 
if we consider that, for example, a popular speaker (i.e., one with many 
neighbors) is given more weight by her interlocutors, for example as in 
\cite{Fagyal2010}. Alternatively, speakers might divide their attention between 
all of their interlocutors.  The voter model described in the Introduction 
(see \cite{Castellano2009} for a review) is an example of such a case: copying 
from a randomly chosen neighbor implies that $H_{ij} \propto 1/k_i$, so that
the combined influence of agent $i$'s neighbors is independent of $i$, no matter
how well-connected she is.

We can access a wide range of models in a systematic way by first supposing 
again that $\alpha_i$ is independent of $i$, say $\alpha = 1$, and further 
assuming that
\begin{equation}
\beta_j = Ak_j^{\sigma}
\end{equation}
for some constants $A$ and $\sigma$. We follow the same procedure as in
Sec.~\ref{sec:independent}. Beginning from Eq.~(\ref{r_fixed_alpha}), we
write down the analog of Eq.~(\ref{j_notequal_i}) and arrive again at the
sum in Eq.~(\ref{sum_on_i}). However, now $\beta^2_i k_i$ is replaced by
$k_i^{2\sigma + 1}$ and $\delta_i$ and $\epsilon_i$ are proportional to 
$k^{\sigma+1}_{i}/N$. Expanding in powers of $k^{\sigma+1}_{i}/N$ one finds
\begin{eqnarray}
r &\approx&\frac{4}{N^2\mu_1\mu^2_{\sigma}}
\left\{\frac{\mu_{\sigma+1}\mu_{2\sigma+1}}
{[2\mu_{\sigma+1}+\mu_{1}A^{-1}]} - \frac{\mu_{3\sigma+2}\mu_{1}A^{-1}}
{N [2\mu_{\sigma+1}+\mu_{1}A^{-1}]^2} \right. \nonumber \\
&-& \left. \frac{2\mu_{4\sigma+3}\mu_{1}A^{-1}}
{N^2 [2\mu_{\sigma+1}+\mu_{1}A^{-1}]^3} - \dots \right\}\;.
\label{r_dep_dist}
\end{eqnarray}
For conventional degree distributions, all the moments tend to $N$-independent 
values as $N$ becomes large, and we have $r \sim 1/N^2$ as usual.

\begin{figure}[htb]
  \includegraphics[width=0.48\textwidth]{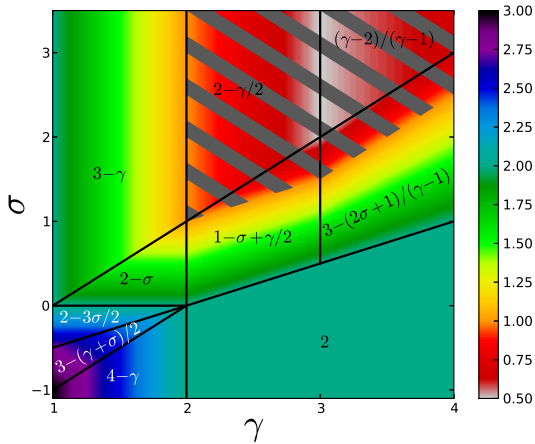}
  \caption{(Color online.) Scaling of mean time to reach fixation with
    population size. Shading represents the value of exponent $\nu$
    where $T \propto N^{\nu}$. The labels give expressions for
    $\nu$ in each region, with black lines marking boundaries
    between regions. The diagonal hatches cover the region in which
    the approximations used are not expected to be accurate.
}\label{exponent_chart}
\end{figure}

Suppose however that the degree distribution obeys a power law.
In different regions of the $\gamma-\sigma$ plane different moments
appearing in (\ref{r_dep_dist}) diverge with $N$. By carefully
examining the ratios between subsequent terms in the series, which
involve ratios of moments $\mu_{(k+1)\sigma + k}/\mu_{k\sigma + k-1}$, we can
establish that in every region the first term dominates. This then leaves 
us with
\begin{equation}\label{simple_r}
r \approx
\frac{4\mu_{\sigma+1}\mu_{2\sigma+1}}{N^2\mu_1\mu_{\sigma}^2[2\mu_{\sigma+1}+
\mu_1A^{-1}]}\;.
\end{equation}
The scaling with respect to $N$ depends on whether any or which
combination of the moments $\mu_1, \mu_{\sigma}, \mu_{\sigma+1},
\mu_{2\sigma+1}$ diverge with $N$ [see
  Eq.~(\ref{powerlaw_scaling})]. This divides the $\sigma$--$\gamma$
plane into a number of regions, as seen in
Fig.~\ref{exponent_chart}. The mean fixation time is proportional to
$1/r$, so finding the population size dependence of
Eq.~(\ref{simple_r}) immediately gives us the scaling of $T$ with
$N$. In general $T\propto N^{\nu}$, and we give expressions for
$\nu$ in the various regions in Fig.~\ref{exponent_chart}.
We see that in a large area, $\nu = 2$ as in the standard case of
$H_{ij}$ all equal. For $\gamma < 3$ and $\sigma < 0$ the mean time to
fixation may grow faster than $N^2$. On the other hand, for $\sigma >
0$ and above the line $\sigma = \gamma-1$, $T$ may grow more slowly
than $N^2$, with the slowest growth rate $T\propto N^{1/2}$ being
achieved when $\gamma = 3$ for $\sigma \geq 2$ (though, as we will
see, our approximations start to break down when $\nu < 1$).

In principle one could also consider further variations, such as $\alpha_i$ 
which are inversely proportional to degree (as in the voter model, or
the uniform listening scenario) and so on. These we  investigate primarily
through numerical simulations, as described below.

\section{Numerical calculations of mean fixation time}
\label{sec:numerical}

To check these calculations, and to explore the robustness of our results when
assumptions we have made are relaxed, we performed Monte Carlo simulations of
the stochastic algorithm described in Sec.~\ref{sec:model}. Explicitly, in
each update, we selected a pair of speakers $i$ and $j$ from the distribution
$G_{ij}$, generated an utterance $\zeta$ for each speaker, and then applied the
update rule (\ref{update_rule}) to both speakers. This update was repeated 
until a state of fixation was reached; the mean time to reach fixation is 
then obtained by averaging over multiple runs. Unless otherwise stated, we 
used homogeneous initial conditions, that is, all $x_i(0)$ are initially set 
to the same value $x_0$.

\subsection{Check of analytical results}

\begin{figure}[t]
  \includegraphics[width=0.48\textwidth]{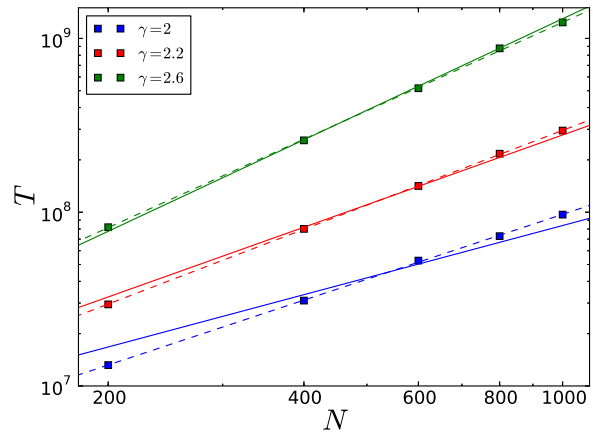}
  \caption{Mean fixation time as a function of population size for
    $H_{ij}$ independent of degree, as described in Sec.
    \ref{sec:independent}. Results are for a fully connected network
    with $\beta_i$ following a power-law distribution with decay
    exponent $\gamma = 2.0,2.2,2.6$ and constant $\alpha_i$. Markers
    are average fixation times for 5000 numerical runs. Solid lines
    are expected scaling as given by Eq.~(\ref{r_largeN}), dashed lines
    are best fit curves of the form $T = AN^{\zeta}$. }\label{TvsN_A}
\end{figure}

We first performed numerical simulations of the situations described in 
Secs.~\ref{sec:independent} and \ref{sec:dependent} to check our results. 
We set $H_{ij} = \alpha_i\beta_j$, and held the $\alpha_i$ values constant.
For the results shown in Fig.~\ref{TvsN_A} we considered a fully
connected network of speakers, that is, each speaker is equally likely
to speak with each of the other speakers, and chose the $\beta_j$ from
a power-law distribution for various values of the power-law exponent
$\gamma$.

We found that the agreement with the predictions of
Eq.~(\ref{r_largeN}) was very good so long as the predicted exponent of
growth of $T$ with $N$ was greater than $1$, that is for $\gamma >
2$. This includes the region $2 < \gamma < 3$, in which the mean
fixation time, $T$, grows more slowly with $N$ than in the usual
situation where $T \propto N^2$. That is, the mean time to fixation
may be reduced without recourse to any special network structure,
merely by allowing heterogeneity in the response of speakers to the
utterances of their interlocutors.

For $\gamma \leq 2$, Eq.~(\ref{r_largeN}) predicts $T \propto N$. As we
we approach this region, we find the theoretical predictions
break down. This can be seen in the lowest set of data in the figure. 
This is not unexpected, if we consider the approximations made to
derive our estimates of the mean fixation time. We have assumed that
there is a short relaxation period after which the dynamics can be
well described by considering only the collective variable
$\xi$ (see \cite{Blythe2010} for details). Our calculated fixation times are
only for this second stage. Typically the initial relaxation happens
in a time of order $N$. We see that if the calculated fixation time is of a 
similar time scale, the initial relaxation can no longer be ignored. This 
is the case whenever $\nu$ approaches $1$ when $T \propto N^{\nu}$.

Similar results were obtained for a sparse interaction network in
which each speaker had approximately an equal number of neighbors.  Thus
shortened fixation times are not a consequence of all agents being able
to interact with all other agents.

\begin{figure}[b]
  \includegraphics[width=0.48\textwidth]{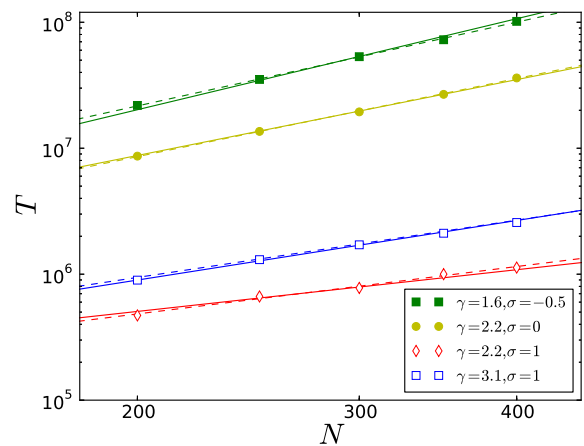}
  \caption{Mean fixation time as a function of population size with
    $\beta_{i}$ depending on speaker degree, as described in Section
    \ref{sec:dependent}. Results are for an random network whose
    degree distribution obeys a power law $p(k) \propto
    k^{-\gamma}$. The interaction weights depend on degree through
    $\beta_i \sim k_i^{\sigma}$. Markers are average fixation times for
    5000 numerical runs. Solid lines are expected scaling as given by
    Eq.~(\ref{simple_r}) and Fig.~\ref{exponent_chart}, dashed lines
    are best fit curves of the form $T = AN^{\zeta}$.}\label{TvsN_B}
\end{figure}

In Fig.~\ref{TvsN_B} we present simulation results for the situation
in which $\beta_i$ does depend on the speaker degree. Specifically,
speakers were placed on an uncorrelated random network whose degree
distribution follows a power law with exponent $\gamma$. These networks 
were generated using the modified configuration model described in 
\cite{Catanzaro2005}. We then set
$\beta_i = k_i^{\sigma}$. The results shown are for various locations
in the $\gamma$--$\sigma$ plane (see Fig.~\ref{exponent_chart}). The
mean fixation time grows with population size as $T \propto
N^{\nu}$ with the value of $\nu$ depending on the parameters
$\gamma$ and $\sigma$. The numerical results are in excellent
agreement with the $\nu$ values predicted by Eq.~(\ref{simple_r}) for
values both smaller and larger than the baseline value of $2$. As
before, we found that the agreement fails when the predicted value of
$\zeta$ is $1$ or less. This occurs in the region marked with diagonal
hatching in Fig. \ref{exponent_chart}.

\subsection{Robustness of the analytical results}

We now discuss cases where the conditions for our analytical results,
Eqs.~(\ref{r_largeN}) and (\ref{simple_r}) do not hold, but we see
nevertheless somewhat similar behavior.

First we investigated the effects of fixed $\beta_i$ and heterogeneous
$\alpha_i$ (on a homogeneous network). By examining
Eq.~(\ref{assume_both}) in this case, we see
that it is the smallest values of $\alpha_i$ which contribute most to
$r$. In fact we find that $r \sim \langle 1/\alpha\rangle /N^2$.  This result
is similar to that found in \cite{Masuda2010, Baxter2011} where different agents in the
network could change state with different rates: this is one way to interpret
variation of the $\alpha$ parameter in the present work.

In this context, we considered a power-law distribution of values, such 
that $P(\alpha) \propto \alpha^{-\gamma}$. The moment $\langle 1/\alpha\rangle$ 
is independent of $\gamma$ in this case [see Eq.~(\ref{y_moment})], so we
would expect to find $T \propto N^2$. Indeed this is exactly what we
observe through numerical simulations, with the mean fixation time
growing as $N^2$, exactly as in the standard case, regardless of the
value of $\gamma$.  

\begin{figure}[tb]
  \includegraphics[width=0.48\textwidth]{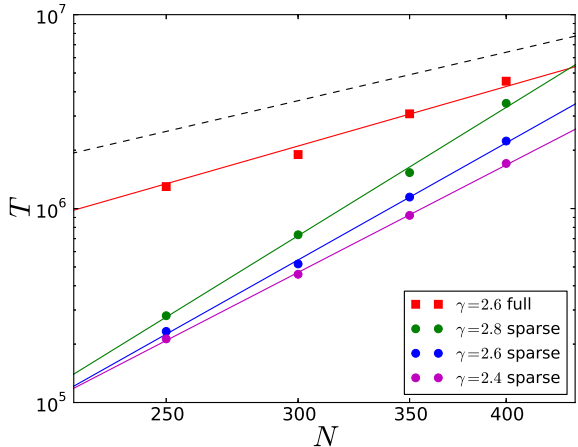}
  \caption{Numerical results for mean fixation time for
    $\alpha_i$ distributed according to
    inverted power law distributions, with values of $\gamma$ given in
    the legend. Top line (squares, red online) is for a fully connected graph
    with $\gamma = 2.6$. Lower lines (circles) are for a sparse graph
    with mean degree $10$ and $\gamma =  2.8, 2.6 , 2.4$ from top to
    bottom. Lines are fitted functions of the form $T =
    aN^{\xi}$. Dashed line is $aN^2$ for comparison.}
\label{inverted_alpha}
\end{figure}

Considering the fact that the smallest $\alpha$ values make the
largest contribution, we also carried out simulations with an
`inverted' power-law distribution, $P(1/\alpha)
\propto (1/\alpha)^{-\gamma}$, that is $P(\alpha)
\propto \alpha^{+\gamma}$ with an imposed upper bound instead of a
lower bound. In this case we do see mean fixation times changing with
$\gamma$, but rather than fixation being sped up, it is slowed
down. We find $T \propto N^{\nu}$, with $\nu$ approaching the baseline
value of $2$ when $\gamma = 3$, and increasing as $\gamma$ decreases,
as shown in Fig. \ref{inverted_alpha}.  Here we do find a difference
relative to other cases we investigated, in that the density of the
graph also has an effect on the exponent $\nu$: it grows more quickly
with decreasing $\gamma$ on a sparse network than a fully connected
network. 

\begin{figure}[tb]
  \includegraphics[width=0.48\textwidth]{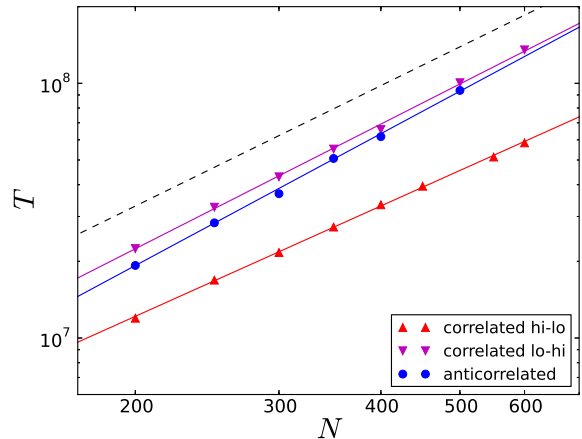}
  \caption{Numerical results for mean fixation time for correlated
    $\beta_i$. Speakers are located on a sparse network and $\beta_i$
    values distributed according to a power law with exponent $\gamma
    = 2.4$, and correlated (see text) from highest to lowest
    (triangles), from lowest to highest (inverted triangles) and
    anti-correlated (circles). Lines are fitted functions of the
    form $T = aN^{\xi}$, with $\xi = 1.44,1.63,1.72$ respectively. For
    comparison the black dashed line has $\xi=1.57$ which is the slope
    expected for uncorrelated $\beta_i$.}\label{correlated_betas}
\end{figure}

Returning to heterogeneous $\beta_i$ values, we investigated the
effect of correlations between the $\beta_i$ values of neighboring
speakers. To do this, we placed the speakers on a random sparse
network, in which each speaker has approximately the same number of
neighbors. A list of power law distributed $\beta$ values was created,
and the largest value assigned to a randomly chosen speaker. The next
largest $\beta$ values were then assigned to the neighbors of this
speaker, followed by remaining second-neighbors and so on until all
$\beta$ values were assigned. We found that these correlations only
slightly affected the scaling of mean fixation time with population
size $N$, with $T$ growing as $N^{\nu}$ with exponent $\nu$ similar to 
that found in Section \ref{sec:independent} for the same $\gamma$. To 
confirm this result, we repeated the simulations now assigning $\beta$ 
values from lowest to highest, and considered anticorrelations, in 
which the lowest $\beta$ values were located on the neighbors of the 
highest value and so on. In each case the growth of $T$ with $N$ was 
similar, though the overall prefactor was different to that found in 
Section \ref{sec:independent}. Results are plotted in 
Fig. \ref{correlated_betas}, compare with Fig. \ref{TvsN_B}. This 
weak dependence of fixation times on correlations mirrors results found 
for the voter model on heterogeneous networks \cite{Sood05}.

Finally we introduced inhomogeneity in the initial conditions. After
randomly assigning $\beta_i$ values exactly as in
Sec.~\ref{sec:independent}, speakers with the largest $\beta_i$'s had
their initial grammar value $x_i(0)$ set to $1$, while the remainder
were set to $0$, such that the overall mean grammar was $x_0$. Our
calculation assumes the largest contribution to
mean fixation time comes from the period after the initial relaxation
to a quasi-stationary state, so the initial conditions would not be
expected to affect the scaling of mean fixation time with $N$. This
was indeed found to be the case, with $T$ scaling with $N$ exactly as
found in Sec.~\ref{sec:independent}. The mean fixation time is
affected by initial conditions through the center-of-mass parameter
$\xi(0)$ which appears in Eq.~(\ref{fix_time}). This affects the
prefactor but not the scaling of $T$ with $N$. We found that $\xi(0)$
differed from the homogeneous case value $x_0$, as evidenced by a much
greater probability of fixation to state $1$.

These last numerical investigations thus support the value of the simpler 
cases for which we made analytic predictions. We find that they give a 
good indication of the general conditions for finding fixation times 
shorter than the standard $T \propto N^2$.

\section{Discussion}
\label{sec:discussion}

In this work, we have investigated how asymmetry in the interactions
between speakers in a model of language change affects the time to
reach a state of fixation (all speakers using a common conventional variant).
Although we have couched our discussion in terms of the utterance selection
model for language change \cite{Baxter06}, it is worth recalling that the 
Fokker-Planck equation that describes the continuous-time limit of the 
dynamics, Eq.~(\ref{FPE}), also applies to the Wright-Fisher model for 
changes in gene frequencies in a structured population \cite{Crow70}, to 
Hubbell's model of ecological community dynamics \cite{Hubbell2001} and, 
in a limit where all $m_{ij} \to 0$, to a spatially-structured voter model 
\cite{Blythe2010}.  Thus our results apply quite generally to models in 
which the state of a node on a network evolves by copying the state of 
a neighboring node, whether through a birth-death process (as in the 
Wright-Fisher or Hubbell model) or by one agent adopting another agent's 
behavior (as in the voter and utterance selection models).

As we noted in the introduction, an appealing and useful property of the
utterance selection model is that there is a clean and natural separation
between the symmetric and asymmetric components of the agent interactions. 
It is assumed that agents'
linguistic behavior is primarily affected by face-to-face interactions
between speakers. Thus whenever agent $i$ is interacting with agent
$j$, agent $j$ is interacting with agent $i$.  This is reflected in
the symmetry of the matrix $G$, viz, $G_{ij} = G_{ji}$. However it is
not necessarily the case that the outcome of the interaction is the
same for both speakers: agent $i$ may be influenced to a greater
degree by agent $j$ than the other way round.  In this instance
$H_{ij} > H_{ji}$, which results in an asymmetric $H$ matrix \footnote{We 
remark that if one extends beyond face-to-face interactions to include 
the case of mass media, for example, this can still be represented with 
a $G_{ij}$ that reflects how often agent $i$ listens to radio station $j$, 
but with a totally asymmetric influence relationship $H_{ij}>0, H_{ji}=0$.}.

This formulation has allowed us to explore in a systematic way the 
consequences of asymmetry in the dynamics by manipulating the $H$ matrix 
while leaving the $G$ matrix unchanged. This is much harder to do in the 
context of the voter model (for example), in which the asymmetry is implicit 
in the model dynamics, rather than specified explicitly as here.  Whilst 
various attempts have been made to separate these two contributions within 
the context of the voter model, see e.g.~\cite{Schneider-Mizell2009,Moretti12},
the network structure and asymmetry effects have generally remained entangled 
to some degree when using the voter model as a starting point.

Our main finding is that the mean time to fixation can be dramatically
affected by the presence of large disparities in the influence of different 
agents, for example, when the $H_{ij}$ are constructed to be drawn from a 
power-law distribution. We emphasize the distinction with similar results 
for the voter model on heterogeneous networks 
(e.g.,~\cite{Sood05,Suchecki05,Suchecki05b,Sood08}), in which
variation in the degree of each node (combined with the implicit
asymmetry of the voter model dynamics) is responsible for such
effects.  Here we find that the fixation time can be reduced relative
to the case of uniform influence ($H_{ij} = {\rm const}$) even on
homogeneous networks.  This result contrasts with those of
\cite{Masuda2010,Baxter2011}, in which variation in the willingness to
change state (our $\alpha$ parameter) causes a slower onset of
fixation, a result we also obtained here.

The specific networks we examined were fully-connected network and
sparsely-connected random graphs.  We have found that, as in earlier
work \cite{Baxter08}, analytical predictions hold when there is a
separation of timescales between an initial relaxation and the longer
diffusive process that brings the system to fixation.  A formal
criterion for this separation of timescales is given in \cite{Blythe2010}, 
but in practice we have found the diffusive timescale dominates when it 
grows more rapidly than linearly with the size of the network $N$. We 
note that this separation of timescales is in fact seen on the 
two-dimensional square lattice (although the diffusive timescale is only 
a factor $\ln N$ longer than the relaxation timescale, \cite{Cox89}).
It is therefore likely that our results hold for the very large class of
networks that satisfy the  `small-world' property, that is, where the
longest distance between any pair of nodes is much smaller than the
network size $N$, not just the random graphs that we considered here
\cite{Newman10}.

We also found that a wide variety of scaling relationships between the 
mean fixation time and network size are possible when node influence 
and degree (a measure of `popularity') co-vary.  Here we found cases 
where fixation may be accelerated or decelerated relative to the baseline 
case of uniform influence, depending on how influence and degree are 
correlated.  Our results are summarized in the phase diagram of 
Fig.~\ref{exponent_chart}, and are similar in spirit to those obtained 
in the specific context of the voter model on heterogeneous 
networks \cite{Schneider-Mizell2009,Moretti12}.

Finally, we find that correlations in influence between neighboring nodes 
only weakly affects the mean time to fixation. This accords for example 
with a similar finding for degree correlations for the voter model on 
heterogeneous networks \cite{Sood05,Blythe2010}, in which correlations 
only appear to affect prefactors in the scaling relation between fixation 
time and network size, not the scaling exponent. This lack of sensitivity 
to correlations may be due once again to the `small-world' property: 
since a variant can reach any node on the network in only a few steps, 
the question of who is using it may become only a second-order
consideration. 

Taken together with the many results for random-copying processes of various 
guises that are to be found in the literature, we have by now a more-or-less 
complete understanding of the factors that enter into the fixation time 
in these models.  There do however remain some generalizations and extensions 
that remain to be fully explored.  Most notably, we have assumed a fixed 
network structure: it is clear that this structure may also evolve over 
time, for example, as  relationships are formed and broken between members 
of a social group.  Furthermore, all the manifestations of the model we 
have discussed share the common and crucial property of neutrality with 
respect to the different variants.  That is, the probability that agent 
$i$ adopts agent $j$'s behavior is independent of what that behavior 
actually is: there is no selection in the language of genetics or ecology.  
While both generalizations have been the subject of considerable study 
(e.g.~\cite{Nardini08,Vazquez08} examine dynamic networks and \cite{Cherry03} 
selection in a spatial setting) the role of network structure and 
interaction asymmetry seems to be less well established in these cases.

\begin{acknowledgments}
GJB thanks the FCT for the support of post-doctoral fellowship
SFRH/BPD/74040/2010. RAB thanks RCUK for the support of an Academic
Fellowship.
\end{acknowledgments}

\bibliography{language}

\begin{appendix}
\section{Moments of Power-law Distributions}

In this paper we frequently write results in terms of moments of 
distributions of network properties. We are often interested in distributions 
with unusually large values, since these model situations where some of the 
speakers have atypical characteristics. In this Appendix we collect together
results on moments of power-law distributions, which are of this kind, that
are used in the main text.

Examples of quantities that we are interested in are: the degree $k_i$ of 
nodes of the network of speakers $G_{ij}$ or the matrix of the weights of 
utterances $H_{ij}$. These are to be sampled from a given distribution. For
generic distributions, the moments are not expected to depend on the sample 
size $N$. However for `heavy-tailed' distributions, the range of values 
likely to be taken by the samples grows with $N$, and as a consequence the 
various moments grow as some power of $N$.

Suppose that the probability distribution of some random variable $q$ takes the 
form
\begin{equation}
P(q) = Aq^{-\gamma} \mbox{ for } q \geq q_0\;,
\label{powerlaw}
\end{equation}
where $A, \gamma$ and $q_0$ are constants. In the limit $N \to \infty$ there
will be arbitrary large values of $q$ which are sampled. In this case the
range of values of $q$ is unbounded ($q_0 < q < \infty$) and simple 
integration gives the normalization constant $A$ as 
$A = (\gamma-1) q_0^{\gamma - 1}$ ($\gamma > 1$) and the $n^{\rm th}$ moment 
$\mu_n$ as
\begin{equation}
\mu_n = \frac{\gamma-1}{\gamma-1-y}q_0^n\;,
\label{yth_moment}
\end{equation}
which diverges if $n > \gamma-1$.

In real applications, and in particular in this paper, we are interested in
the case where $N$ is finite. In this case we expect that there will be some 
upper cutoff $q_{\rm max}$ that grows with $N$. The easiest way to extract the 
scaling of this cutoff with $N$ is to put
\begin{equation}
\int_{q_{\rm max}}^{\infty} q^{-\gamma} \propto \frac{1}{N} \;,
\end{equation}
motivated by the requirement that the values of $q$ not seen due to finite 
sample-size effects will have a cumulative probability of order $1/N$.  
Performing the integral and rearranging yields $q_{\rm max} \sim N^{1/
(\gamma-1)}$ (see e.g.~\cite{Sood05}).  More rigorously, one can compute the 
distribution of the maximum of $N$ power-law random numbers, which has the
Fr\'echet form
\begin{equation}
P_N(q) \sim  N (\gamma-1) q^{-\gamma} {\rm e}^{-N q^{1-\gamma}}
\end{equation}
for large $N$ and $q_0=1$ (see e.g.~\cite{Castillo2005}).
Using this distribution, one can now calculate the mean value 
of the maximum $q$ for a given $N$, which is found to scale in the same 
way as before, $q_{\rm max} \sim N^{1/(\gamma-1)}$.  In the context of networks, 
however, there is an additional condition, in that we do not wish to have any 
multiple edges.  This yields the upper cutoff $\propto N^{1/2}$ for 
$\gamma < 3$~\cite{Boguna2004}. Setting $q_{\rm max} = aN^{1/\rho}$ with 
$\rho = 2$ for $\gamma \leq 3$ and $\rho = (\gamma-1)$ for $\gamma > 3$, 
leads to
\begin{equation}
\mu_n = \frac{\gamma-1}{\gamma-1-n}\frac{[q_0^{1-\gamma+n} - 
a^{1-\gamma+n}N^{(1-\gamma+n)/\rho}]}{[q_0^{1-\gamma} - 
a^{1-\gamma}N^{(1-\gamma)/\rho}]}\;.
\label{y_moment}
\end{equation}
If $n < \gamma-1$, the terms in Eq.~(\ref{y_moment}) containing $N$ decay with 
increasing $N$, leading to a value for the moment (for sufficiently large $N$) 
close to that found in the case of infinite $N$. On the other hand, when 
$n > \gamma-1$, but $\gamma > 1$, the term in $N$ in the numerator diverges, 
while that in the denominator vanishes, leaving
\begin{equation}
\mu_n \approx \frac{\gamma-1}{n-(\gamma-1)}
\frac{a^{1-\gamma+n}N^{(1-\gamma+n)/\rho}}{q_0^{1-\gamma}}\;.
\end{equation}
In summary, for $\gamma >1$, the $n^{\rm th}$ moment is of order:
\begin{eqnarray}
\mu_n \sim \begin{cases}
N^{(1-\gamma+n)/2} & n > \gamma -1 \qquad \gamma \leq 3\\
N^{(1-\gamma+n)/(\gamma-1)} & n > \gamma -1 \qquad \gamma > 3\\
q_0^n \sim 1 & n < \gamma - 1\;.\label{powerlaw_scaling}
\end{cases}
\end{eqnarray}

\end{appendix}

\end{document}